\documentclass{appolb}
\usepackage{graphicx}
\usepackage{amsmath,amssymb}
\usepackage{epstopdf}

\newcommand{\mbf}[1]{\mathbf{#1}}
\renewcommand{\bar}[1]{\overline{#1}}

\usepackage[usenames]{color}
\usepackage[colorlinks=true, urlcolor=navyblue, linkcolor=navyblue, citecolor=navyblue]{hyperref}
\usepackage{relsize}
\definecolor{navyblue}{rgb}{0,0.08,0.45}

\begin{document}
\title{QCD and Light-Front Holography%
\thanks{Presented by SJB at  Light Cone 2012: 8-13 July, 2012, Polish Academy of Arts and Sciences,  
 Cracow, Poland}%
}
\author{Stanley J. Brodsky
\address{ SLAC National Accelerator Laboratory\\
Stanford University, Stanford, CA 94309, USA}
{
$ $
\\
$ $
}
\\Guy F. de T\'eramond \\
\address{ Universidad de Costa Rica, San Jos\'e, Costa Rica}  
}   
\maketitle
\begin{abstract}
The eigenvalues of the light-front  QCD Hamiltonian, quantized  at fixed light-front  time $\tau  = t+z/c$,  predict the hadronic mass spectrum and the corresponding eigensolutions provide the light-front wavefunctions which describe hadron structure. 
More generally, we show that the valence Fock-state wavefunctions 
of the light-front QCD Hamiltonian satisfy a single-variable relativistic equation of motion, analogous to the nonrelativistic radial Schr\"odinger equation, with an effective 
confining potential $U$ which systematically incorporates the effects of higher quark and gluon Fock states.   We outline a method for computing the required potential from first principles in QCD.
The holographic mapping of gravity in AdS space to QCD, quantized at fixed light-front time, yields the same light front Schr\"odinger equation;  in fact,  the soft-wall AdS/QCD approach  provides  a model for the light-front potential which is color-confining and reproduces well the light-hadron spectrum.  One also derives via light-front holography a precise relation between the bound-state amplitudes in the fifth dimension of AdS space and the boost-invariant light-front wavefunctions describing the internal structure of hadrons in physical space-time.   
The elastic and transition form factors of the pion and the nucleons are found to be well described in this framework.  
The light-front AdS/QCD  holographic approach thus gives  a frame-independent first approximation of the color-confining dynamics,  spectroscopy, and excitation spectra of relativistic light-quark bound states in QCD.    
\end{abstract}

\section{Introduction to Light-Front Hamiltonian QCD}

A central goal of hadron and nuclear physics is to compute the spectrum and dynamics of hadrons directly from the QCD Lagrangian, as illustrated in Fig. 1.   Lattice gauge theory is  a traditional method. However, another important nonperturbative method for solving QCD is quantization at fixed light-front time $\tau = x^+ = t+z/c$  -- Dirac's ``Front Form".~\cite{Dirac:1949cp} Unlike lattice gauge theory, the solutions are in Minkowski space, there is no fermion doubling problem, and the formalism is frame-independent.  
Solving a quantum field theory is then equivalent to determining the eigensolutions of the light-front (LF) Hamiltonian -- the relativistic field theoretic analog of the Heisenberg equation. 

\begin{figure}[h]
\centering
\includegraphics[width=8cm]{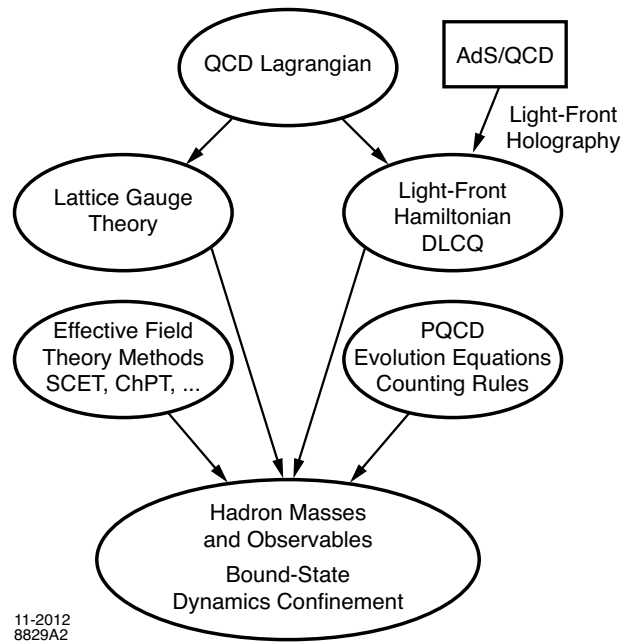}  
 \caption{Methods for solving  QCD. Light-front holography provides a model for the color-confining potential in the QCD LF Hamiltonian.}
\label{NPQCD}
\end{figure} 

Remarkably,  one can use ``Light-front holography" to show that the AdS/QCD theory based on the dynamics of a gravitational theory in the 
fifth dimension of  anti de-Sitter (AdS) space has a direct mapping to QCD defined at fixed light-front time.~\cite{deTeramond:2008ht}  In fact, the light-front 
AdS/QCD  holographic approach provides a frame-independent first approximation to the color-confining dynamics,  spectroscopy, and excitation spectra of relativistic light-quark bound states in QCD.   

As emphasized by Dirac, the quantization of a relativistic quantum field theory at fixed light-front   time  has the extraordinary advantage that 
boost transformations are kinematical rather than dynamical.~\cite{Dirac:1949cp}
Hadronic eigenstates $P_\mu P^\mu  = M^2$ are determined by the
Lorentz-invariant Hamiltonian equation for the relativistic bound-state 
\begin{equation} \label{LFH}
H_{LF} \vert  \psi(P) \rangle =  M^2 \vert  \psi(P) \rangle,
\end{equation}
with  $H_{LF} \equiv P_\mu P^\mu  =  P^- P^+ -  \mbf{P}_\perp^2$.   The LF Hamiltonian $H_{LF}$  defined at fixed $\tau$ is a Lorentz scalar. One can derive  the LF Hamiltonian directly from the QCD Lagrangian using canonical methods.~\cite{Brodsky:1997de} If one chooses light-cone gauge, the quanta of the gluon field have physical polarization $S^z = \pm 1$ and no ghost degrees of freedom. 
Solving nonperturbative QCD is thus equivalent to solving the Heisenberg matrix eigenvalue problem.   Angular momentum $J^z$ is conserved at every vertex.
The LF vacuum is defined as the state of lowest invariant mass and is trivial up to zero modes.  There are thus no quark or gluon vacuum condensates in the LF vacuum-- the corresponding physics is contained within the 
LFWFs themselves~\cite{Brodsky:2009zd,Brodsky:2010xf}, thus eliminating a major contribution to the cosmological constant.

The hadronic state $\vert\psi\rangle$ is an expansion in multiparticle Fock states
$\vert \psi \rangle = \sum_n \psi_n \vert n \rangle$, where the components $\psi_n = \langle n \vert \psi \rangle$ are a column vector of states, and the basis vectors $\vert n \rangle$ are the $n$-parton eigenstates of the free LF Hamiltonian: 
$\vert q \bar q \rangle , \vert q \bar q g \rangle,   \vert q \bar q  q \bar q \rangle,  \cdots$ etc.  
This Fock space expansion of the hadronic eigensolutions of the LF Hamiltonian 
provides the light-front wavefunctions (LFWFs) $\psi_{n/H}(x_i, \vec k_{\perp i}, \lambda_i) $ which define the hadron's structure in the same sense that the Schr\"odinger wavefunction in nonrelativistic quantum mechanics describes the physics of atoms.  Remarkably, the LFWFs are independent of the hadron's total four momentum  $P^\mu$ and thus of the observer's Lorentz frame;
no boosts are required.   
The Fock state expansion begins with the dominant valence state  ({\it e.g.}, $ n =3 $ for the nucleon); higher Fock states have extra gluons and $q \bar q$ pairs. The constituents satisfy $x_i = k^+_i/P^+$, $\sum^n_{i=1} x_i$,  $\sum^n_{i=1} \vec k_{\perp i} = 0$.  The $\lambda_i$ specify their spin projections in the $\hat z$ direction, as in non-relativistic physics.  Each Fock state satisfies $J^z$ conservation.

Hadronic observables can be computed directly from the LFWFs: structure functions 
and transverse momentum distributions (TMDs) are computed from the absolute squares of the LFWFs; in fact, LFWFs are the basis of the parton model.   Elastic and transition form factors, decay amplitudes, and generalized parton distributions are computed from overlaps of the LFWFs.
A key example of the utility of the light-front formalism is the Drell-Yan West formula~\cite{Drell:1969km,West:1970av} for the space-like form factors of electromagnetic currents given as overlaps of initial and final LFWFs. 
The valence and sea quark and gluon
distributions which are measured in deep inelastic lepton scattering
are defined from the squares of the LFWFs summed over all Fock
states $n$. 
The gauge-invariant ``distribution
amplitudes" $\phi_H(x_i,Q)$ defined from the integral over the
transverse momenta $\mbf{k}^2_{\perp i} \le Q^2$ of the valence
(smallest $n$) Fock state provide a fundamental measure of the
hadron at the amplitude level;~\cite{Lepage:1979zb,Efremov:1979qk}
they  are the nonperturbative inputs to the factorized form of hard
exclusive amplitudes and exclusive heavy hadron decays in perturbative
QCD.  

At high momentum where one can iterate the hard scattering kernel, one can derive dimensional counting rules~\cite{Brodsky:1973kr,Matveev:1973ra}
and factorization theorems.
The resulting structure functions distributions obey  DGLAP  evolution equations, and the distribution amplitudes obey 
ERBL evolution equations as a function of the maximal invariant
mass, thus providing a physical factorization
scheme.~\cite{Lepage:1980fj} In each case, the derived quantities
satisfy the appropriate operator product expansions, sum rules, and
evolution equations. At large $x$ where the struck quark is
far-off shell, DGLAP evolution is quenched,~\cite{Brodsky:1979qm} so
that the fall-off of the DIS cross sections in $Q^2$ satisfies Bloom-Gilman
inclusive-exclusive duality at fixed $W^2.$

Given the light-front wavefunctions $\psi_{n/H}$, one can
compute a large range of other hadron
observables. Exclusive weak transition
amplitudes~\cite{Brodsky:1998hn} such as $B\to \ell \nu \pi$,  and
the generalized parton distributions~\cite{Brodsky:2000xy} measured
in deeply virtual Compton scattering  $\gamma^* p \to \gamma p$ are (assuming the ``handbag"
approximation) overlap of the initial and final LFWFs with $n
=n^\prime$ and $n =n^\prime+2$.  
It is also possible to compute jet hadronization at the amplitude level from first principles from the LFWFs.~\cite{Brodsky:2008tk} A similar method has been used to predict the production of antihydrogen from the off-shell coalescence of relativistic antiprotons and positrons.~\cite{Munger:1993kq}

The simplicity of the front form contrasts with the usual instant-form formalism. Current matrix elements defined at ordinary time $t$ must include the coupling of the photons and vector bosons fields  to connected vacuum currents; otherwise the result is not Lorentz-invariant.  Thus the knowledge of the hadronic eigensolutions of the instant-form Hamiltonian are insufficient for determining form factors or other observables.   In addition, the boost of an instant form wavefunction from $p$ to $p+q$ changes particle number and is an extraordinarily complicated dynamical problem.

In some cases such as the $T$-odd Sivers  single-spin asymmetry in semi-inclusive deep inelastic scattering,~\cite{Brodsky:2002cx} Drell-Yan lepton-pair 
production,~\cite{Collins:2002kn,Brodsky:2002rv} and single-particle inclusive reactions, one must include the ``lensing" (Wilson-line) effects of initial or final-state interactions.~\cite{Brodsky:2010vs} The ordinary rules of factorization are broken since the sign of the single spin asymmetry depends on whether one has initial-state or final-state lensing.~\cite{Collins:2002kn,Brodsky:2002rv}   Diffractive deep inelastic scattering at leading twist is also due to the color-neutralization of final state lensing.  Double-initial state interactions break the Lam-Tung relation at  leading twist.~\cite{Boer:2002ju}
The LFWFs of hadrons thus provide a direct connection between observables and the QCD Lagrangian. They are as essential to hadron physics as DNA is to biology!

Theories such as $QCD(1+1) $ can be solved to any desired accuracy for any number of colors, flavors, and quark masses by diagonalization of the Heisenberg LF matrix by employing periodic boundary conditions as in the discretized light-cone quantization method (DLCQ).~\cite{Pauli:1985ps,Hornbostel:1988fb}  Solving QCD(3+1) using  DLCQ is much more challenging since the matrix diagonalization yields all hadronic eigensolutions simultaneously.  Vary {\it et al.}~\cite{Vary:2009gt} are developing an alternative method which uses a basis of orthonormal harmonic states generated by AdS/QCD.   A LF coupled cluster  method for solving quantum field theories has been developed by Chabysheva and Hiller.~\cite{Chabysheva:2011ed}
We will discuss in the next section a new method where one can solve for  each hadron eigenstate  in terms of a single-variable LF equation analogous to the radial Schr\"odinger equation in atomic physics.  
We shall also show that, remarkably, one obtains a single-variable equation of the same form with a color-confining potential using AdS/QCD and LF 
Holography.~\cite{deTeramond:2008ht}  In all of these LF methods, the observer's frame is arbitrary.

Other advantages of light-front quantization include:

\begin{itemize}

\item
The simple structure of the light-front vacuum allows an unambiguous
definition of the partonic content of a hadron in QCD.  
The chiral and gluonic condensates can be identified with the  hadronic LF Fock states,~\cite{Brodsky:2009zd,Casher:1974xd} rather than vacuum properties.  LF zero mode contributions are also possibly relevant.~\cite{Banks:1979yr}  
As shown by a Bethe Salpeter analysis,~\cite{Maris:1997tm} the usual Gell-Mann-Oakes-Renner (GMOR) relation~\cite{GellMann:1968rz} is satisfied in QCD, but with the vacuum to vacuum matrix element $\langle 0|q \bar q|0 \rangle $ replaced by  $\langle 0|q \gamma_5|\pi \rangle/ f_\pi$.
This matrix element requires a pion $q \bar q$ LF  wavefunction  with
$L^z= 1, S^z=-1$  or    $L^z= -1, S^z=+1$, which occurs when a quark mass insertion causes a quark chiral flip.  
For a recent review see Ref.~\cite{Brodsky:2012ku}.

\item
Since it is defined along the front of a light-wave, at fixed $\tau$, the LF front-form vacuum is causal, and it is thus 
a match to the empty universe with zero cosmological constant. In contrast, the instant-form vacuum describes a state of indeterminate energy at fixed time $t$, a  boundary condition outside the causal horizon.
The QCD LF vacuum is trivial up to zero modes in the front form, thus eliminating contributions to the cosmological constant.~\cite{Brodsky:2009zd} 
In the case of the Higgs model, the effect of the usual Higgs vacuum expectation value is replaced by a constant $k^+=0$ zero mode field.~\cite{Srivastava:2002mw}  

\item
If one quantizes QCD in the physical light-cone gauge (LCG) $A^+ =0$, then the gluons have physical angular momentum projections $S^z= \pm 1$. The orbital angular momenta of quarks and gluons are defined unambiguously, and there are no ghosts.

\item
The gauge-invariant distribution amplitude $\phi(x,Q)$  is the integral of the valence LFWF in LCG integrated over the internal transverse momentum $k^2_\perp < Q^2$ because the Wilson line is trivial in this gauge. It is also possible to quantize QCD in  Feynman gauge in the light front.~\cite{Srivastava:1999gi}

\item
LF Hamiltonian perturbation theory provides a simple method for deriving analytic forms for the analog of Parke-Taylor amplitudes~\cite{Motyka:2009gi} where each particle spin $S^z$ is quantized in the LF $z$ direction.  The gluonic $g^6$ amplitude  $T(-1 -1 \to +1 +1 +1 +1 +1 +1)$  requires $\Delta L^z =8;$ it thus must vanish at tree level since each three-gluon vertex has  $\Delta L^z = \pm 1.$ However, the order $g^8$ one-loop amplitude can be nonzero.

\item
Amplitudes in light-front perturbation theory may be automatically renormalized using the ``alternate denominator"  subtraction method. The application to QED has been checked at one and two loops.~\cite{Brodsky:1973kb}

\item
A fundamental theorem for gravity can be derived from the equivalence principle:  the anomalous gravitomagnetic moment defined from the spin-flip  matrix element of the energy-momentum tensor is identically zero, $B(0)=0$.~\cite{Teryaev:1999su} This theorem can be proven in  the light-front formalism Fock state by Fock state.~\cite{Brodsky:2000ii} 

\item
LFWFs obey the cluster decomposition theorem, providing an elegant proof of this theorem for relativistic bound states.~\cite{Brodsky:1985gs}

\item
LF quantization provides a distinction~\cite{Brodsky:2009dv} between static  (the square of LFWFs) distributions versus non-universal dynamic structure functions,  such as the Sivers single-spin correlation and diffractive deep inelastic scattering which involve final state interactions.  The origin of nuclear shadowing and process-independent anti-shadowing also becomes explicit.~\cite{Brodsky:2004qa}

\item
LF quantization provides a simple method to implement jet hadronization at the amplitude level.~\cite{Brodsky:2008tk}

\item
The instantaneous fermion interaction in LF  quantization provides a simple derivation of the $J=0$
fixed pole contribution to deeply virtual Compton scattering,~\cite{Brodsky:2009bp} i.e., the $e^2_q s^0 F(t)$  contribution to the DVCS amplitude which is independent of photon energy and virtuality.

\item
Unlike instant time quantization, the bound state
Hamiltonian equation of motion in the LF is frame independent. This makes a direct connection of QCD with AdS/CFT methods possible.~\cite{deTeramond:2008ht}

\end{itemize}

\section{The Light-Front Schr\"odinger Equation: a Semiclassical Approximation to QCD \label{LFQCD}}

It is greatly advantageous to reduce the full multiparticle eigenvalue problem of the LF Hamiltonian to an effective light-front Schr\"odinger equation (LFSE) which acts on the valence sector LF wavefunction and determines each eigensolution separately.~\cite{Pauli:1998tf}   In contrast,  diagonalizing the LF Hamiltonian yields all eigensolutions simultaneously,
a complex task.
The central problem for deriving the LFSE becomes the derivation of the effective interaction $U$ which acts only on the valence sector of the theory and has, by definition, the same eigenvalue spectrum as the initial Hamiltonian problem. For carrying out this program one most systematically express the higher Fock components as functionals of the lower ones. The method has the advantage that the Fock space is not truncated and the symmetries of the Lagrangian are preserved.~\cite{Pauli:1998tf}

For example, we can derive the  LFSE for a meson  in terms of the invariant impact-space variable for a two-parton state $\zeta^2= x(1-x)\mbf{b}_\perp^2$ 
\begin{equation} \label{eq:psiphi}
\psi(x,\zeta, \varphi) = e^{i L \varphi} X(x) \frac{\phi(\zeta)}{\sqrt{2 \pi \zeta}} ,
\end{equation}
thus factoring the angular dependence $\varphi$ and the longitudinal, $X(x)$, and transverse mode $\phi(\zeta)$.
In the limit of zero quark masses the longitudinal mode decouples and
the LF eigenvalue equation $P_\mu P^\mu \vert \phi \rangle  =  M^2 \vert \phi \rangle$
is thus a light-front  wave equation for $\phi$~\cite{deTeramond:2008ht}
\begin{equation} \label{LFWE}
\left[-\frac{d^2}{d\zeta^2}
- \frac{1 - 4L^2}{4\zeta^2} + U\left(\zeta^2, J, M^2\right) \right]
\phi_{J,L,n}(\zeta^2) = M^2 \phi_{J,L,n}(\zeta^2),
\end{equation}
a relativistic {\it single-variable}  LF  Schr\"odinger equation. 

The potential in the LFSE is determined from the two-particle irreducible (2PI) $ q \bar q \to q \bar q $ Greens' function.  In particular, the higher Fock states in intermediate states
leads to an effective interaction $U(\zeta^2, J ,M^2)$  for the valence $\vert q \bar q \rangle$ Fock state.~\cite{Pauli:1998tf}
The potential $U$ thus depends on the hadronic eigenvalue $M^2$ via the LF energy denominators 
$P^-_{\rm initial} - P^-_{\rm intermediate} + i \epsilon$
of the intermediate states which connect different LF Fock states.  
Here $P^-_{\rm initial} =( {M^2 + \mathbf{P}^2_\perp)/ P^+}$. The dependence of $U$ on $M^2$ is analogous to the retardation effect in QED interactions, such as the hyperfine splitting in muonium, which involves the exchange of a propagating photon.  Accordingly, the eigenvalues $M^2$ must be determined 
self-consistently.~\cite{deTeramond:2012cs} 
The  $M^2$ dependence of the effective potential thus reflects the contributions from higher Fock states in the LFSE (\ref{LFWE}),  since
 $U(\zeta^2, J ,M^2)$ is also the kernel for the  scattering amplitude $q \bar q \to q \bar q$ at $s = M^2.$    It has only ``proper'' contributions; {\it i.e.}, it has no $q \bar q$ intermediate state.  The potential can be constructed, in principle systematically, using LF time-ordered perturbation theory.  In fact, as we shall see below, the QCD theory has the identical form as the AdS theory, but with the quantum field-theoretic corrections due to the higher Fock states giving a general form for the potential.  This provides a novel way to solve nonperturbative QCD.  The LFSE for QCD becomes increasingly accurate as one includes contributions from very high particle number Fock states.

The above discussion assumes massless quarks. More generally we must include mass terms~\cite{Brodsky:2008pg,Gutsche:2011uj}
${m^2_a / x}  + {m^2_b/(1-x)}$ in the kinetic energy term and allow  the potential $  U(\zeta^2, x, J,M^2)$ to have dependence on the LF momentum fraction $x$.  The quark masses also appear in $U$ due to the  presence in the LF denominators as well as the chirality-violating interactions connecting the valence Fock state to the higher Fock states. In this case, however, the equation of motion cannot be reduced to a single variable.

The LFSE approach also can be applied to atomic bound states in QED and nuclei. In principle one could compute the spectrum and dynamics of atoms, such as the Lamb shift and hyperfine splitting of hydrogenic atoms to high precision by a systematic treatment of the potential. Unlike the ordinary  instant form, the resulting LFWFs are independent of the total momentum and can thus describe ``flying atoms'' without the need for dynamical boosts, 
such as the ``true muonium'' $(\mu^+ \mu^-)$ bound states which can be produced  by Bethe-Heitler pair production $\gamma \, Z \to (\mu^+ \mu^-)  \, Z$  below 
threshold.~\cite{Brodsky:2009gx} A related approach for determining the valence light-front wavefunction and studying the effects of higher Fock states without truncation has been given in Ref.~\cite{Chabysheva:2011ed}.

\section{Effective Confinement Interaction from the Gauge/gravity Correspondence}

A remarkable correspondence between the equations of motion in AdS  and the Hamiltonian equation for relativistic bound-states  was found in Ref.~\cite {deTeramond:2008ht}.   In fact, to a first semiclassical approximation,
LF QCD  is formally equivalent to the equations of motion on a fixed gravitational background~\cite{deTeramond:2008ht} asymptotic to AdS$_5$, where  confinement properties  are encoded in a dilaton profile $\varphi(z)$.

A spin-$J$ field in AdS$_{d+1}$ is represented by a rank $J$ tensor field $\Phi_{M_1 \cdots M_J}$, which is totally symmetric in all its indices.  In presence of a dilaton background field $\varphi(z)$ the action is~\footnote{The study of   higher integer and half-integer spin wave equations  in  AdS  is based on our collaboration with Hans Guenter Dosch.~\cite{DBT:2012}  See also the discussion in Ref.~\cite{Gutsche:2011vb}.}
\begin{multline} \label{SJ}
S = \frac{1}{2} \int \! d^d x \, dz  \,\sqrt{g} \,e^{\varphi(z)} 
  \Big( g^{N N'} g^{M_1 M'_1} \cdots g^{M_J M'_J} D_N \Phi_{M_1 \cdots M_J} D_{N'}  \Phi_{M'_1 \cdots M'_J}    \\
 - \mu^2  g^{M_1 M'_1} \cdots g^{M_J M'_J} \Phi_{M_1 \cdots M_J} \Phi_{M'_1 \cdots M'_J}  + \cdots \Big)  ,
\end{multline}
where $M, N = 1, \cdots , d+1$, $\sqrt{g} = (R/z)^{d+1}$ and $D_M$ is the covariant derivative which includes parallel transport. 
The coordinates of AdS are the Minkowski coordinates $x^\mu$ and the holographic variable $z$, $x^M = \left(x^\mu, z\right)$. The  d + 1 dimensional mass $\mu$ is not a physical observable and is {\it a priory} an arbitrary
parameter. The dilaton background field $\varphi(z)$ in  (\ref{SJ})   introduces an energy scale in the five-dimensional AdS action, thus breaking its conformal invariance. It  vanishes in the conformal ultraviolet limit $z \to 0$.

 A physical hadron has plane-wave solutions and polarization indices along the 3 + 1 physical coordinates
 $\Phi_P(x,z)_{\mu_1 \cdots \mu_J} = e^{- i P \cdot x} \Phi(z)_{\mu_1 \cdots \mu_J}$,
 with four-momentum $P_\mu$ and  invariant hadronic mass  $P_\mu P^\mu \! = M^2$. All other components vanish identically. 
 One can then construct an effective action in terms
 of the spin modes $\Phi_J = \Phi_{\mu_1 \mu_2 \cdots \mu_J}$ with only  physical degrees of 
 freedom. In this case the system of coupled differential equations which follow from (\ref{SJ}) reduce to a homogeneous equation in terms of the physical field $\Phi_J$
  upon rescaling the AdS mass $\mu$
 \begin{equation} \label{AdSWEJ}
\left[-\frac{ z^{d-1 -2 J}}{e^{\varphi(z)}}   \partial_z \left(\frac{e^\varphi(z)}{z^{d-1 - 2 J}} \partial_z\right)
+ \left(\frac{\mu R}{z}\right)^2\right] \Phi(z)_J = M^2 \Phi(z)_J .
 \end{equation}

Upon the substitution $z \! \to\! \zeta$  and 
$\phi_J(\zeta)   = \left(\zeta/R\right)^{-3/2 + J} e^{\varphi(z)/2} \, \Phi_J(\zeta)$ 
in (\ref{AdSWEJ}), we find for $d=4$ the LFSE (\ref{LFWE}) with effective potential~\cite{deTeramond:2010ge}
\begin{equation} \label{U}
U(\zeta^2, J) = \frac{1}{2}\varphi''(\zeta^2) +\frac{1}{4} \varphi'(\zeta^2)^2  + \frac{2J - 3}{2 \zeta} \varphi'(\zeta^2) ,
\end{equation}
provided that the fifth dimensional mass $\mu$ is related to the internal orbital angular momentum $L = max \vert L^z \vert$ and the total angular momentum $J^z = L^z + S^z$ according to $(\mu R)^2 = - (2-J)^2 + L^2$.  The critical value  $L=0$  corresponds to the lowest possible stable solution, the ground state of the LF Hamiltonian.
For $J = 0$ the five dimensional mass $\mu$
 is related to the orbital  momentum of the hadronic bound state by
 $(\mu R)^2 = - 4 + L^2$ and thus  $(\mu R)^2 \ge - 4$. The quantum mechanical stability condition $L^2 \ge 0$ is thus equivalent to the Breitenlohner-Freedman stability bound in AdS.~\cite{Breitenlohner:1982jf}
The scaling (twist)  dimensions are $2 + L$ independent of $J$, in agreement with the twist-scaling dimension of a two-parton bound state in QCD.  

The correspondence between the LF and AdS equations  thus determines the effective confining interaction $U$ in terms of the infrared behavior of AdS space and gives the holographic variable $z$ a kinematical interpretation. The identification of the orbital angular momentum 
is also a key element of our description of the internal structure of hadrons using holographic principles.

\begin{figure}[h]
\centering
\includegraphics[width=6.15cm]{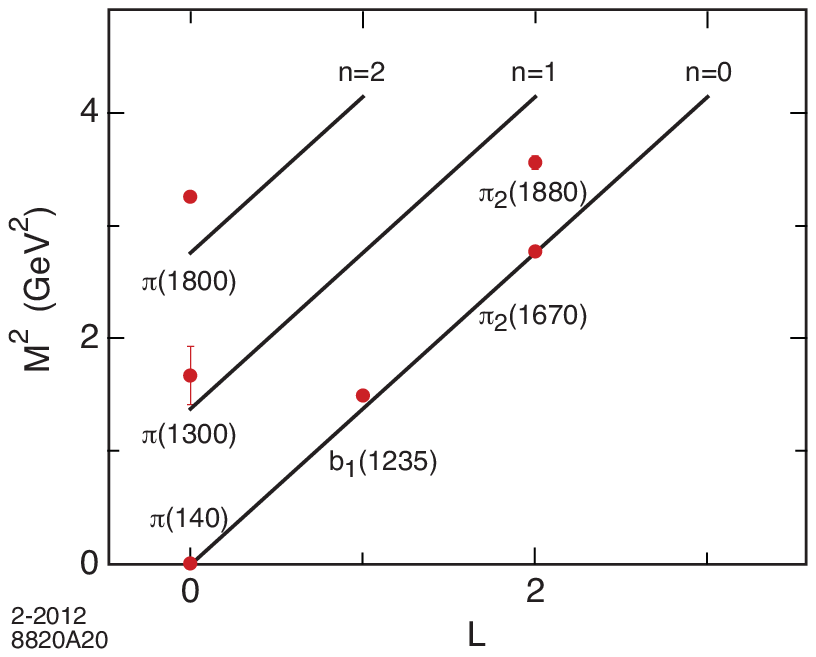}  \hspace{0pt}
\includegraphics[width=6.15cm]{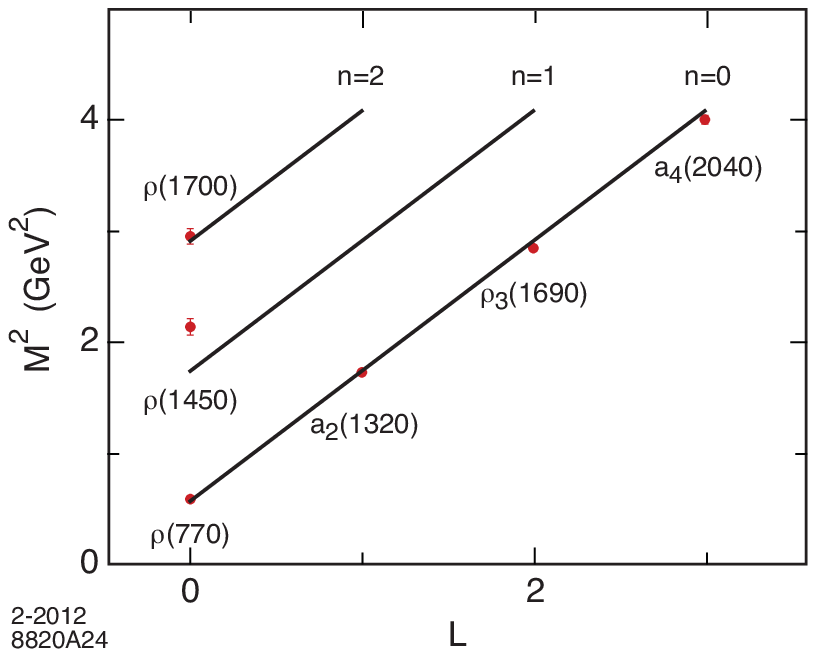}
 \caption{$I = 1$ parent and daughter Regge trajectories for the $\pi$-meson family (left) with
$\kappa= 0.59$ GeV; and  the   $\rho$-meson
 family (right) with $\kappa= 0.54$ GeV.}
\label{pionspec}
\end{figure} 

\begin{figure}[h]
\begin{center}
 \includegraphics[width=6.15cm]{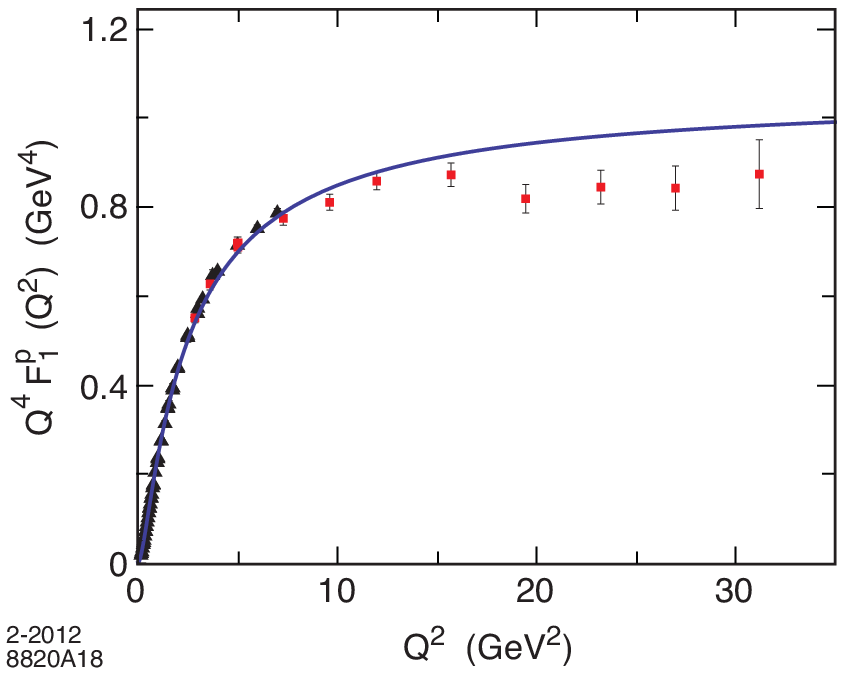}   \hspace{0pt}
\includegraphics[width=6.15cm]{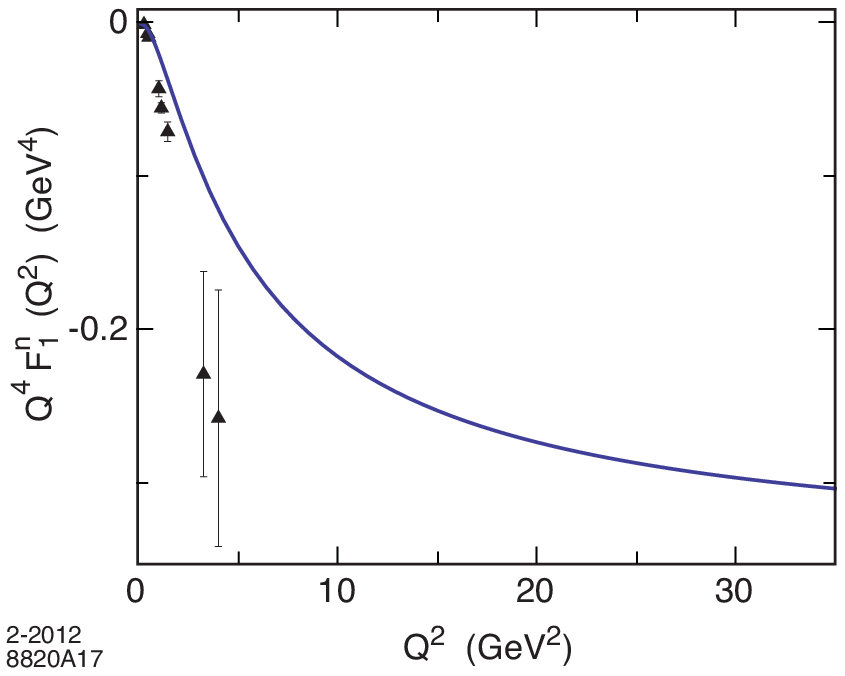}
 \includegraphics[width=6.15cm]{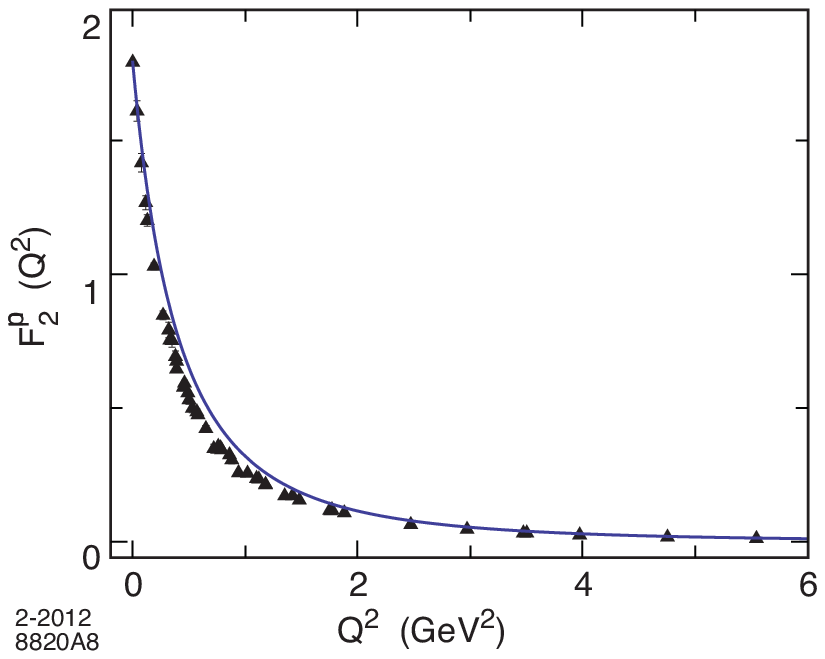}   \hspace{0pt}
\includegraphics[width=6.15cm]{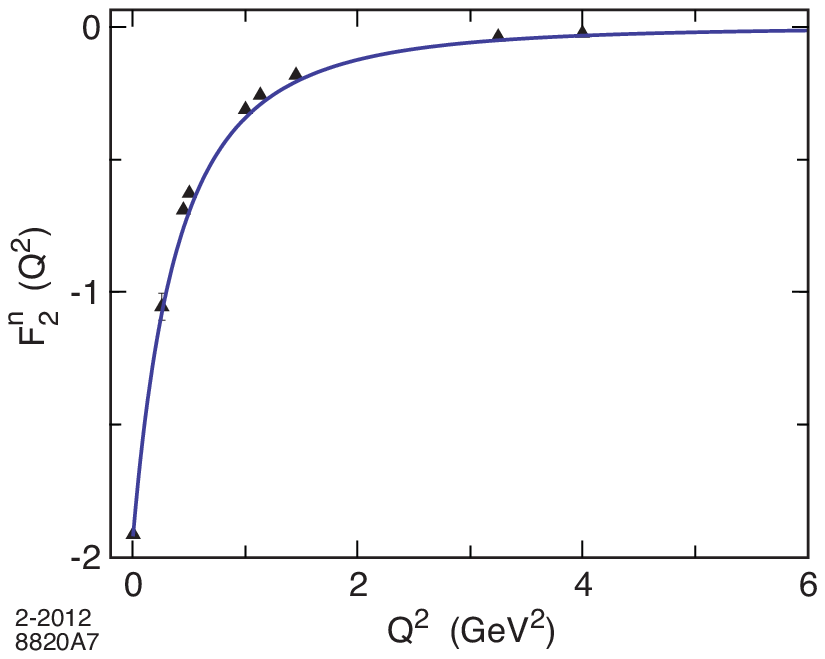}
 \caption{Predictions for  $Q^4 F_1^p(Q^2)$ (upper left) and   $Q^4 F_1^n(Q^2)$ (upper right) in the
soft wall model. The results for  $F_2^p(Q^2)$ (lower left)  and  $F_2^n(Q^2)$ (lower right) are normalized to static quantities. 
The nucleon spin-flavor structure corresponds to the $SU(6)$ limit.}
\label{fig:nucleonFF}
\end{center}
\end{figure}

A particularly interesting example is a dilaton profile $\exp{\left(\pm \kappa^2 z^2\right)}$ of either sign, since it 
leads to linear Regge trajectories~\cite{Karch:2006pv} and avoids the ambiguities in the choice of boundary conditions at the infrared wall.  
For the  confining solution $\varphi = \exp{\left(\kappa^2 z^2\right)}$ the effective potential is
$U(\zeta^2,J) =   \kappa^4 \zeta^2 + 2 \kappa^2(J - 1)$ and  Eq.  (\ref{LFWE}) has eigenvalues
$M_{n, J, L}^2 = 4 \kappa^2 \left(n + \frac{J+L}{2} \right)$,
with a string Regge form $M^2 \sim n + L$.  
A discussion of the light meson and baryon spectrum,  as well as  the elastic and transition form factors of the light hadrons using LF holographic methods, is given in 
Ref.~\cite{deTeramond:2012rt}.  As an example the spectral predictions  for the $J = L + S$ light pseudoscalar and vector meson  states are  compared with experimental data in Fig. \ref{pionspec} for the positive sign dilaton model.  The data is from PDG.~\cite{PDG2012}   The results for $Q^4 F_1^p(Q^2)$ and $Q^4 F_1^n(Q^2)$  are shown in Fig. \ref{fig:nucleonFF}.  We also show in  Fig. \ref{fig:nucleonFF}
the results for $F_2^p(Q^2)$ and $F_2^n(Q^2)$ for the same value of $\kappa$ normalized  to the static quantities $\chi_p$ and $\chi_n$.
To compare with physical data we have shifted the poles appearing in the expression of the form factor to their physical values located at $M^2 = 4 \kappa^2(n + 1/2)$ 
following the  discussion in Ref.~\cite{deTeramond:2012rt}.
The value $\kappa = 0.55$ GeV  is determined from the $\rho$ mass.  The data compilation  is from Ref.~\cite{Diehl:2005wq}.

Holographic QCD also provide powerful nonperturbative analytical tools to extend the multi-component light-front Fock structure of amplitudes to the time-like region.
Consider for example the elastic pion form factor $F_\pi$. The pion is a superposition of an infinite number of Fock components  $\vert N \rangle$, $\vert \pi\rangle = \sum_N \psi_N \vert N \rangle$, and thus 
the pion form factor is  given by~\cite{deTeramond:2012rt}
$
F_\pi(s) = \sum_\tau P_\tau F_\tau(s),
$
where $P_\tau$ is the probability for the twist $\tau$ (the number of components $N$ in a given Fock-state) and  $F_\tau$
\begin{equation}   \label{Ftau} 
F_\tau(s) =  \frac{M^2_{n=0} M^2_{n=1} \cdots M^2_{n=\tau-2} }{{\left(M^2_{n=0} - s \right) }
 \left(M^2_{n=1} - s\right)  \cdots 
       \left(M^2_{n=\tau-2}  - s\right)}.
\end{equation}   
is  written as a $\tau - 1$ product of poles.~\cite{Brodsky:2007hb}  Normalization at $Q^2 =0$,  $F_\pi(0) = 1$, implies that 
 $\sum_\tau P_\tau = 1$ if all possible states are included.
 
 In the strongly coupled semiclassical gauge/gravity limit, hadrons have zero widths and are stable,  as  in the  $N_C \to \infty$ limit of QCD.  One can nonetheless modify  (\ref{Ftau})  by introducing  finite widths  using the substitution 
 $s \to s + i \sqrt{s} \, \Gamma$. The resulting expression  has a series of resonance poles  in the non-physical lower-half complex $ \sqrt{s}$-plane (Im $\sqrt{s} <0$) located at  $\alpha^n_\pm = \frac{1}{2}\left(- i \Gamma_n \pm \sqrt{4 M_n^2 - \Gamma_n^2 } \right)$ with $n = 0, 1, \cdots, \tau-2$. In the upper half plane  (Im $\sqrt{s} >0$) --which maps to the physical sheet, the pion form factor is an analytic function and thus can be mapped to the space-like region by a $\pi/2$ rotation in the complex $\sqrt{s}$-plane such as to avoid cutting the real axis. This simple procedure allows us to analytically continue the holographic results to the space-like region where the pion form factor is also modified   by the finite widths of the vector meson resonances from interactions with the continuum.~\cite{Arriola:2012vk}

\begin{figure*}[h]
\centering 
\includegraphics[width=12.6cm]{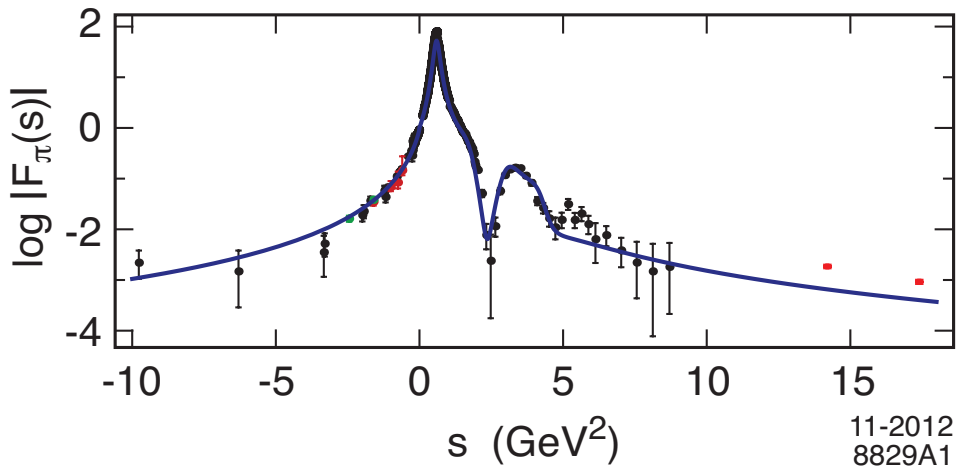}  
\caption{Structure of the space-like ($s<0$)  and time-like ($s> 4 m_\pi^2$)  pion  form factor in light-front holography for a truncation of the pion wave function up to twist five.
The space-like data are  the compilation  from Baldini  {\it et al.}~\cite{Baldini:1998qn} (black)  and JLAB data~\cite{Tadevosyan:2007yd} (green and red). The time-like data
are from the recent precise measurements from BABAR~\cite{Aubert:2009ad} (black)  and CLEO~\cite{Seth:2012nn} (red).}
\label{pionFF}
\end{figure*} 

To illustrate the relevance of higher Fock states in the analytic  structure of the pion form factor we consider a simple phenomenological model where we include the first three components in a Fock expansion of the pion state $\vert \pi \rangle  = \psi^{L=0}_{q \bar q /\pi} \vert q \bar q; L=0   \rangle_{\tau=2} 
+  \psi^{L=0}_{q \bar q q \bar q} \vert q \bar q  q \bar q;  L=0  \rangle_{\tau=4}  +  \psi^{L=1}_{q \bar q q \bar q} \vert q \bar q  q \bar q; L =1\rangle_{\tau=5} + \cdots,$ and no constituent dynamical gluons.~\cite{Brodsky:2011pw}  The $J^{PC} = 0^{- +}$ twist-2, twist-4 and twist-5 states  are created by the interpolating operators $\mathcal{O}_2 = \bar q \gamma^+ \gamma_5  q$,  $\mathcal{O}_4 = \bar q \gamma^+ \gamma_5  q  \bar q q$ and $\mathcal{O}_5 = \bar q \gamma^+ \gamma_5  D^\mu q  \bar q \gamma_\mu q$. The results for the space-like and time-like form factor are shown in Fig. \ref{pionFF}. We choose the values $\Gamma_\rho =  149$ MeV,   $\Gamma_{\rho'} =  300$ MeV,  $\Gamma_{\rho''} =  250$ MeV and $\Gamma_{\rho'''} = 200$ MeV. The chosen width of the $\rho'$ is slightly smaller than the PDG value listed in Ref. \cite{PDG2012}. The widths of the $\rho$ and $\rho''$ are the PDG  values and the width of the $\rho'''$ is unknown. 
 The results correspond to $P^{L=0}_{q \bar q q \bar q}$ = 2.5 \% and  $P^{L=1}_{q \bar q q \bar q}$ = 4.5 \% the admixture of the
$\vert q \bar q q \bar q; L=0  \rangle$ and $\vert q \bar q q \bar q; L=1  \rangle$ states. The value of $P^{L=0}_{q \bar q q \bar q}$ and $P^{L=1} _{q \bar q q \bar q}$ (and the widths) are input in the model. The value of $\kappa= 0.55$ GeV is determined from the $\rho$ mass and the masses of the radial excitations follow from setting the poles at their physical locations, $M^2 \to 4 \kappa^2(n + 1/2)$, as discussed in~\cite{deTeramond:2012rt}. The main features of the space-like and time-like regions are well described by the same formula with a minimal number of parameters. In contrast, models  based on hadronic degrees of freedom involve sums over a large number of intermediate states and thus require, for large $s$, a very large number of hadronic parameters~\cite{Kuhn:1990ad, Bruch:2004py, Hanhart:2012wi}.  A detailed data comparison requires the introduction  of thresholds and $s$-dependent widths from multiparticle states coupled to the $\rho$-resonances. This will be described elsewhere.~\cite{dTB}

Despite some limitations of AdS/QCD, the LF holographic  approach to the gauge/gravity duality,  has  given significant physical  insight into the strongly-coupled nature and internal structure of hadrons.  In particular, the AdS/QCD soft-wall model provides an elegant analytic framework for describing  nonperturbative  hadron dynamics, the systematics of the excitation spectrum of hadrons, including their empirical multiplicities and degeneracies. It also provides powerful new analytical tools for computing hadronic transition amplitudes incorporating conformal scaling behavior at short distances and the transition from the hard-scattering perturbative domain, where quark and gluons are the relevant degrees of freedom, to the long-range confining hadronic region.

The effective interaction $ U(\zeta^2) = \kappa^4 \zeta^2 + 2 \kappa^2(J-1)$  that is derived from the AdS/QCD model is instantaneous in LF time and acts on the lowest state of the LF Hamiltonian.  This equation describes the spectrum of mesons as a function of $n$, the number of nodes in $\zeta^2$ and the total angular momentum $J=J^z$,
with $J^z = L^z + S^z$  the sum of the internal orbital angular momentum of the constituents~\footnote{The  $SO(2)$ Casimir  $L^2$  corresponds to  the group of rotations in the transverse LF plane.} and their internal spin.
It is the relativistic frame-independent front-form analog of the non-relativistic radial Schr\"odinger equation for muonium  and other hydrogenic atoms in presence of an instantaneous Coulomb potential.

The AdS/QCD harmonic oscillator potential could in fact emerge from the exact QCD formulation when one includes contributions from the LFSE potential $U$ which are due to the exchange of two connected gluons; {\it i.e.}, ``H'' diagrams.~\cite{Appelquist:1977tw}
We notice that $U$ becomes complex for an excited state since a denominator can vanish; this gives a complex eigenvalue and the decay width.

\section{Future Applications}

The  AdS/QCD model can be considered as a first approximation to the full QCD dynamics of hadrons.  The analytic form of the LFWFs obtained by solving the LFSE for meson and baryons can be utilized for many fundamental issues in hadron physics  and can be tested experimentally for a wide range of observables.
These include 
\begin{itemize}

\item
Timelike hadron form factors using the multipole structure of the dressed current defined by AdS/QCD

\item 
Hadronization at the amplitude level~\cite{Brodsky:2008tk} 

\item 
Resolving the factorization uncertainty in DGLAP evolution and fragmentation

\item Application of holographic LFWFs to diffractive $\rho$ meson electroproduction~\cite{Forshaw:2012im,Dosch:2006kz} and diffractive physics

\item Resonant structure in photon-photon amplitudes such as the $J^{PC} = 0^{++}$ channel in $\gamma \gamma \to \pi \omega$~\cite{JD}

\item Behavior of the QCD running coupling at soft scales~\cite{Brodsky:2010ur}

\item The computation of transition form factors such as  photon-to-meson form factors~\cite{Brodsky:2011xx} and heavy-hadron weak decays

\item The physics of higher Fock states including intrinsic heavy quark excitations~\cite{Chang:2011du}

\item Hidden-color degrees of freedom in nuclear wavefunctions~\cite{Brodsky:1983vf}

\item Color transparency phenomena~\cite{Brodsky:1988xz,ElFassi:2012nr} 

\item  The sublimation of gluon degrees of freedom at low virtuality~\cite{Brodsky:2011pw}

\item  The shape of valence structure functions, distribution amplitudes, and generalized parton distributions~\cite{Vega:2012iz}

\item The emergence of dimensional counting rules at $x \to 1$ in inclusive reactions and hard exclusive reactions~\cite{Brodsky:1994kg}

\item New insights into the breaking of chiral symmetry~\cite{Kapusta:2011zza}

\item Renormalization group properties under conformal transformations~\cite{Glazek:2012am}

\end{itemize}

\section{Summary}

We have shown that the valence Fock-state wavefunctions of the eigensolutions of the light-front QCD Hamiltonian satisfy a single-variable relativistic equation of motion, analogous to the nonrelativistic radial Schr\"odinger equation with an effective 
confining potential $U$, which systematically incorporates the effects of higher quark and gluon Fock states.   We have outlined a method for computing the required potential from first principles. 
The holographic mapping of gravity in AdS space to QCD, quantized at fixed light-front time, yields the same LF Schr\"odinger equation;  in fact,  
the soft-wall  AdS/QCD approach  provides  a model for the LF potential which is color-confining and reproduces well the light-hadron spectrum.  One also derives via light-front holography a precise relation between the bound-state amplitudes in the fifth dimension of AdS space and the boost-invariant light-front wavefunctions describing the internal structure of hadrons in physical space-time.   
The elastic and transition form factors of the pion and the nucleons are well described in this framework.  
The light-front AdS/QCD  holographic approach thus gives  a frame-independent first approximation of the color-confining dynamics,  spectroscopy, and excitation spectra of relativistic light-quark bound states in QCD.    This provides  the basis for a profound connection between physical QCD, quantized on the light-front, and the physics of hadronic modes in a higher dimensional AdS space.

\section*{Acknowledgments}

We thank Rinaldo Baldini, Michel Davier, Joseph Day, Hans Guenter Dosch, Sean Dobbs, Simon Eydelman, Jeff Forshaw, Stanislaw Glazek, Siegfried Krewald,  Stefan Mueller,  Michael Peskin, Ruben Sandapen and Kamal Seth for helpful comments. 
This research was supported by the Department of Energy contract DE--AC02--76SF00515.
Invited talk, presented by SJB at  Light Cone 2012: 8-13 July, 2012, Polish Academy of Arts and Sciences,  Cracow, Poland. SLAC-PUB-15288.

\end{document}